\documentclass{article}
\usepackage{graphicx}
\usepackage{amsmath}

\input{tcilatex}

\begin{document}

\title{\textbf{Noncommutative Two Time Physics }}
\author{W. Chagas-Filho \\
Departamento de Fisica, Universidade Federal de Sergipe\\
SE, Brazil}
\maketitle

\begin{abstract}
We present a classical formalism describing two-time physics with Abelian
canonical gauge field backgrounds. The formalism can be used as a starting
point for the construction of an interacting quantized two-time physics
theory in a noncommutative space-time.
\end{abstract}

\section{Introduction}

Two-Time Physics [1,2,3,4,5,6,7] is an approach that provides a new
perspective for understanding ordinary one-time dynamics from a higher
dimensional, more unified point of view including two time-like dimensions.
This is achieved by introducing a new gauge symmetry that insure unitarity,
causality and absence of ghosts. The new phenomenon in two-time physics is
that the gauge symmetry can be used to obtain various one-time dynamical
systems from the same simple action of two-time physics, through gauge
fixing, thus uncovering a new layer of unification through higher dimensions
[7].

An approach to the introduction of background gravitational and gauge fields
in two-time physics was first presented in [7]. In [7], the linear
realization of the $Sp(2,R)$ gauge algebra of two-time physics is required
to be preserved when background gravitational and gauge fields come into
play. To satisfy this requirement, the background gravitational field must
satisfy a homothety condition [7], while in the absence of gravitational
fields the gauge field must satisfy certain conditions [7] which were first
proposed by Dirac [8] in 1936. Dirac proposed these conditions as subsidiary
conditions to describe the usual 4-dimensional Maxwell theory of
electromagnetism (in the Lorentz gauge) as a theory in 6 dimensions which
automatically displays $SO(4,2)$ symmetry. In the treatment of [7] the gauge
field $A_{M}$ and the gravitational field $G_{MN}$ are explicit functions of
position only.

Explicit dependence on position only for the gravitational and gauge fields
may in some cases be interpreted as a certain restriction on the formalism,
since two-time physics treats $X$ and $P$ as indistinguishable variables. In
the most general situation, the gravitational and gauge fields in two-time
physics must be described by a bi-local gravitational field $G_{MN}(X,P)$
and by a doublet of bi-local gauge fields, $A_{i}^{M}(X,P)$, $i=1,2$, as
discussed in [2] and briefly reviewed in section three below. In this letter
we follow an intermediate path between [2] and [7] and present a formalism
for introducing a single bi-local background gauge field $A_{M}(X,P)$ in
two-time physics. Our formalism reproduces an extended canonical version of
Dirac's first subsidiary condition found in [7,8], and can be used to
construct an interacting quantized two-time physics theory in a
noncommutative space-time. This last observation opens new research
directions in two-time physics.

The paper is divided as follows. In the next section we review the basic
formalism of two-time physics and show how the $SO(d,2)$ Lorentz generator
for the free action can be obtained from a local scale invariance of the
Hamiltonian. The presence of this local scale invariance implies that the
free two-time physics theory can also be consistently formulated in terms of
another set of phase space brackets, with the only difference being that the
linear realization of the gauge algebra is replaced by a non-linear
realization. This replacement does not introduce any inconsistencies into
the formalism because in two-time physics the metric signature with two
timelike dimensions is a requirement that comes from the constraint
equations only, and not from a particular realization of the gauge algebra.
We will see in section two that the linearity or nonlinearity of a gauge
algebra depends on the fundamental set of canonical brackets that are being
used to compute the algebra. In section three we extend these results to the
interacting theory. Some concluding remarks appear in section four.

\section{Two-time Physics}

The central idea in two-time physics [1,2,3,4,5,6,7] is to introduce a new
gauge invariance in phase space by gauging the duality of the quantum
commutator $[X_{M},P_{N}]=i\delta _{MN}$. This procedure leads to a
symplectic Sp(2,R) gauge theory. To remove the distinction between position
and momenta we set $X_{1}^{M}=X^{M}$ and $X_{2}^{M}=P^{M}$ and define the
doublet $X_{i}^{M}=(X_{1}^{M},X_{2}^{M})$. The local Sp(2,R) acts as 
\begin{equation}
\delta X_{i}^{M}(\tau )=\epsilon _{ik}\omega ^{kl}(\tau )X_{l}^{M}(\tau ) 
\tag{2.1}
\end{equation}
$\omega ^{ij}(\tau )$ is a symmetric matrix containing three local
parameters and $\epsilon _{ij}$ is the Levi-Civita symbol that serves to
raise or lower indices. The Sp(2,R) gauge field $A^{ij}$ is symmetric in $%
(i,j)$ and transforms as 
\begin{equation}
\delta A^{ij}=\partial _{\tau }\omega ^{ij}+\omega ^{ik}\epsilon
_{kl}A^{lj}+\omega ^{jk}\epsilon _{kl}A^{il}  \tag{2.2}
\end{equation}
The covariant derivative is 
\begin{equation}
D_{\tau }X_{i}^{M}=\partial _{\tau }X_{i}^{M}-\epsilon _{ik}A^{kl}X_{l}^{M} 
\tag{2.3}
\end{equation}
An action invariant under the Sp(2,R) gauge symmetry is 
\begin{equation}
S=\frac{1}{2}\int d\tau (D_{\tau }X_{i}^{M})\epsilon ^{ij}X_{j}^{N}\eta _{MN}
\tag{2.4a}
\end{equation}
After an integration by parts this action can be written as 
\begin{equation*}
S=\int d\tau (\partial _{\tau }X_{1}^{M}X_{2}^{N}-\frac{1}{2}%
A^{ij}X_{i}^{M}X_{j}^{N})\eta _{MN}
\end{equation*}
\begin{equation}
=\int d\tau \lbrack \dot{X}.P-(\frac{1}{2}\lambda _{1}P^{2}+\lambda _{2}P.X+%
\frac{1}{2}\lambda _{3}X^{2})]  \tag{2.4b}
\end{equation}
where $A^{11}=\lambda _{3}$, $A^{12}=A^{21}=\lambda _{2}$, \ $A^{22}=\lambda
_{1}$ and the canonical Hamiltonian is 
\begin{equation}
H=\frac{1}{2}\lambda _{1}P^{2}+\lambda _{2}P.X+\frac{1}{2}\lambda _{3}X^{2} 
\tag{2.5}
\end{equation}
The equations of motion for the $\lambda $'s give the primary constraints 
\begin{equation}
\phi _{1}=\frac{1}{2}P^{2}\approx 0  \tag{2.6}
\end{equation}
\begin{equation}
\phi _{2}=P.X\approx 0  \tag{2.7}
\end{equation}
\begin{equation}
\phi _{3}=\frac{1}{2}X^{2}\approx 0  \tag{2.8}
\end{equation}
and therefore we can not solve for the $\lambda $'s from their equations of
motion. The values of the $\lambda $'s in action (2.4b) are arbitrary.
Constraints (2.6)-(2.8), as well as evidences of two-time physics, were
independently obtained in [9]. The notation $\approx $ means that
constraints (2.6)-(2.8) ''weakly vanish'' [10]. Therefore, following Dirac's
convention [10] for systems with first-class constraints only, they are set
strongly equal to zero only after all calculations have been performed.

If we consider the Euclidean, or the Minkowski metric as the background
space-time, we find that the surface defined by the constraint equations
(2.6)-(2.8) is trivial. The only metric giving a non-trivial surface and
avoiding the ghost problem is the flat metric with two timelike dimensions
[1,2,3,4,5,6,7]. We then must work in a $(d+2)$ dimensional Euclidean
space-time. We emphasize here that this transition to a $d+2$ dimensional
space-time is an imposition of the constraint equations (2.6)-(2.8).

We use the Poisson brackets 
\begin{equation}
\{P_{M},P_{N}\}=0  \tag{2.9a}
\end{equation}
\begin{equation}
\{X_{M},P_{N}\}=\delta _{MN}  \tag{2.9b}
\end{equation}
\begin{equation}
\{X_{M},X_{N}\}=0  \tag{2.9c}
\end{equation}
where $M,N=1,...,d+2$, and verify that constraints (2.6)-(2.8) obey the
algebra 
\begin{equation}
\{\phi _{1},\phi _{2}\}=-2\phi _{1}  \tag{2.10a}
\end{equation}
\begin{equation}
\{\phi _{1},\phi _{3}\}=-\phi _{2}  \tag{2.10b}
\end{equation}
\begin{equation}
\{\phi _{2},\phi _{3}\}=-2\phi _{3}  \tag{2.10c}
\end{equation}
These equations show that all constraints $\phi $ are first-class
constraints [10]. Equations (2.10) represent the linear symplectic Sp(2,R)
gauge algebra.

Action (2.4) also has a global symmetry under Lorentz transformations $%
SO(d,2)$ with generator [1,2,3,4,5,6,7] 
\begin{equation}
L_{{}}^{MN}=\epsilon ^{ij}X_{i}^{M}X_{j}^{N}=X_{M}P_{N}-X_{N}P_{M} 
\tag{2.11}
\end{equation}
It satisfies 
\begin{equation}
\{L_{MN},L_{RS}\}=\delta _{MR}L_{NS}+\delta _{NS}L_{MR}-\delta
_{MS}L_{NR}-\delta _{NR}L_{MS}  \tag{2.12}
\end{equation}
and is gauge invariant because it has identically vanishing brackets with
the first-class constraints (2.6)- (2.8), $\{L_{MN},\phi _{i}\}=0,$ $%
i=1,2,3. $ In [7], the form (2.11) for the Lorentz generator of two-time
physics is preserved when a background gauge field $A_{M}(X)$ is introduced.
In this letter we will show how, in a phase space with the usual Poisson
brackets, the linear realization (2.10) of the gauge algebra and the form
(2.11) for the Lorentz generator are preserved when a massless Abelian
bi-local background gauge field $A_{M}(X,P)$ is introduced. We will also
show how, in a phase space with Snyder [11] brackets, when the same gauge
field is introduced, the form (2.11) for $L_{MN}$ is preserved but the gauge
algebra acquires a non-linear realization.

Hamiltonian (2.5) is invariant under the scale transformations 
\begin{equation}
X^{M}\rightarrow \tilde{X}^{M}=\exp \{\beta \}X^{M}  \tag{2.13a}
\end{equation}
\begin{equation}
P_{M}\rightarrow \tilde{P}_{M}=\exp \{-\beta \}P_{M}  \tag{2.13b}
\end{equation}
\begin{equation}
\lambda _{1}\rightarrow \exp \{2\beta \}\lambda _{1}  \tag{2.13c}
\end{equation}
\begin{equation}
\lambda _{2}\rightarrow \lambda _{2}  \tag{2.13d}
\end{equation}
\begin{equation}
\lambda _{3}\rightarrow \exp \{-2\beta \}\lambda _{3}  \tag{2.13e}
\end{equation}
where $\beta $ is an arbitrary function of $X$ and $P$. Keeping only the
linear terms in $\beta $ in the transformation (2.13), after some algebra we
arrive at the brackets 
\begin{equation}
\{\tilde{P}_{M},\tilde{P}_{N}\}=(\beta -1)[\{P_{M},\beta \}P_{N}+\{\beta
,P_{N}\}P_{M}]+\{\beta ,\beta \}P_{M}P_{N}  \tag{2.14a}
\end{equation}
\begin{equation*}
\{\tilde{X}_{M},\tilde{P}_{N}\}=(1+\beta )[\delta _{MN}(1-\beta
)-\{X_{M},\beta \}P_{N}]
\end{equation*}
\begin{equation}
+(1-\beta )X_{M}\{\beta ,P_{N}\}-X_{M}X_{N}\{\beta ,\beta \}  \tag{2.14b}
\end{equation}
\begin{equation}
\{\tilde{X}_{M},\tilde{X}_{N}\}=(1+\beta )[X_{M}\{\beta
,X_{N}\}-X_{N}\{\beta ,X_{M}\}]+X_{M}X_{N}\{\beta ,\beta \}  \tag{2.14c}
\end{equation}
for the transformed canonical variables. If we choose $\beta =\phi _{1}=%
\frac{1}{2}P^{2}$ $\approx 0$ in equations (2.14), compute the brackets on
the right hand sides, and impose the vanishing of $\phi _{1}$ at the end of
the calculation, we arrive at the brackets 
\begin{equation}
\{\tilde{P}_{M},\tilde{P}_{N}\}=0  \tag{2.15a}
\end{equation}
\begin{equation}
\{\tilde{X}_{M},\tilde{P}_{N}\}=\delta _{MN}-P_{M}P_{N}  \tag{2.15b}
\end{equation}
\begin{equation}
\{\tilde{X}_{M},\tilde{X}_{N}\}=-(X_{M}P_{N}-X_{N}P_{M})  \tag{2.15c}
\end{equation}
Now, using again $\beta =\phi _{1}=\frac{1}{2}P^{2}\approx 0$ in
transformations (2.13a) and (2.13b) and imposing the vanishing of $\phi _{1}$%
, these become the identity transformations 
\begin{equation}
\tilde{X}_{M}=X_{M}  \tag{2.16a}
\end{equation}
\begin{equation}
\tilde{P}_{M}=P_{M}  \tag{2.16b}
\end{equation}
Substituting equations (2.16) in (2.15) we arrive at the brackets

\begin{equation}
\{P_{M},P_{N}\}=0  \tag{2.17a}
\end{equation}
\begin{equation}
\{X_{M},P_{N}\}=\delta _{MN}-P_{M}P_{N}  \tag{2.17b}
\end{equation}
\begin{equation}
\{X_{M},X_{N}\}=-(X_{M}P_{N}-X_{N}P_{M})  \tag{2.17c}
\end{equation}
Brackets (2.17) are the classical equivalent of the Snyder commutators [11]
which were proposed in 1947 as a way to solve the ultraviolet divergence
problem in quantum field theory. In the canonical quantization procedure,
where brackets are replaced by commutators according to the rule 
\begin{equation*}
\lbrack commutator]=i\{bracket\}
\end{equation*}
the brackets (2.17) will lead directly to a noncommutative quantized [11] $%
d+2$ dimensional space-time for two-time physics.

The classical Snyder brackets (2.17) are associated to a non-linear
realization of the gauge algebra of two-time physics. Computing the algebra
of constraints (2.6)-(2.8) using these brackets we arrive at the expressions 
\begin{equation}
\{\phi _{1},\phi _{2}\}=-2\phi _{1}+4\phi _{1}^{2}  \tag{2.18a}
\end{equation}
\begin{equation}
\{\phi _{1},\phi _{3}\}=-\phi _{2}+2\phi _{1}\phi _{2}  \tag{2.18b}
\end{equation}
\begin{equation}
\{\phi _{2},\phi _{3}\}=-2\phi _{3}+\phi _{2}^{2}  \tag{2.18c}
\end{equation}
Since the space-time metric with two timelike dimensions is a consequence of
the constraint equations (2.6)-(2.8) only, the above non-linear gauge
algebra exactly corresponds to the same expression (2.11) for the $SO(d,2)$
generator. In fact, $L_{MN}$ explicitly appears with a minus sign in the
right hand side of the Snyder bracket (2.17c), giving an interesting
connection of the Lorentz invariance of action (2.4) with the scale
invariance (2.13) of Hamiltonian (2.5).

The conclusion at this point is that the free two-time physics theory can
also be consistently formulated in a phase space where the Snyder brackets
(2.17) are valid. The only difference in this alternative formulation is
that the linear realization (2.10) of the gauge algebra is substituted by
the non-linear realization (2.18). In the next section we will see that this
remains true when an Abelian massless bi-local gauge field $A_{M}(X,P)$ that
satisfies extended Dirac's subsidiary conditions is introduced.

\section{2T Physics with Abelian Gauge Fields}

In two-time physics, interactions with gravitational fields $G_{MN}(X,P)$
and gauge fields $A_{i}^{M}(X,P)$ in a way that respects the $Sp(2,R)$ gauge
symmetry is also possible. In the presence of these interactions the free
action (2.4a) is modified as [2] 
\begin{equation*}
S_{G,A}=\frac{1}{2}\int d\tau \lbrack (D_{\tau }X_{i}^{M})\epsilon
^{ij}X_{j}^{N}G_{MN}(X_{1},X_{2})
\end{equation*}
\begin{equation}
+(D_{\tau }X_{i}^{M})\epsilon ^{ij}A_{jM}(X_{1},X_{2})]  \tag{3.1}
\end{equation}
$G_{MN}$ is a scalar under $Sp(2,R)$ and a symmetric traceless tensor in $%
d+2 $ dimensions. $A_{i}^{M}$ is a doublet under $Sp(2,R)$ and a vector in $%
d+2$ dimensions. For the local $Sp(2,R)$ invariance to hold, there must be
restrictions on the functional forms of both $G_{MN}(X_{1},X_{2})$ and $%
A_{i}^{M}(X_{1},X_{2})$ since the arguments $(X_{1},X_{2})$ also transform
under $Sp(2,R)$. For consistency with local symmetry, gravity and gauge
interactions are more conveniently expressed in terms of bi-local fields $%
G_{MN}(X_{1},X_{2})$ and $A_{i}^{M}(X_{1},X_{2})$ in $d+2$ dimensions [2].
Bi-local fields were advocated in [12,13] as a means of extending
supergravity and super Yang-Mills theory to (10,2) dimensions based on clues
from the BPS solutions of extended supersymmetry. The use of bi-local
gravitational and gauge fields in two-time physics was, however, apparently
not further motivated beyond [2,12,13].

In this letter we study another possibility of introducing background gauge
fields in two-time physics. Here we study the problem of introducing a
single bi-local background gauge field $A_{M}(X_{1},X_{2})$ in two-time
physics. We are able to obtain the necessary conditions for the local $%
Sp(2,R)$ and global $SO(d,2)$ symmetries to hold in this case. In our
treatment, when only one background bi-local gauge field $A_{M}(X(\tau
),P(\tau ))$ is introduced, the free 2T action (2.4b) becomes 
\begin{equation}
S=\int d\tau \{\dot{X}.P-[\frac{1}{2}\lambda _{1}(P-A)^{2}+\lambda
_{2}(P-A).X+\frac{1}{2}\lambda _{3}X^{2}]\}  \tag{3.2}
\end{equation}
where the Hamiltonian is 
\begin{equation}
H=\frac{1}{2}\lambda _{1}(P-A)^{2}+\lambda _{2}(P-A).X+\frac{1}{2}\lambda
_{3}X^{2}  \tag{3.3}
\end{equation}
The equations of motion for the multipliers now give the constraints 
\begin{equation}
\phi _{1}=\frac{1}{2}(P-A)^{2}\approx 0  \tag{3.4}
\end{equation}
\begin{equation}
\phi _{2}=(P-A).X\approx 0  \tag{3.5}
\end{equation}
\begin{equation}
\phi _{3}=\frac{1}{2}X^{2}\approx 0  \tag{3.6}
\end{equation}
We must now define a set of brackets between the canonical variables and the
gauge field. A convenient set is 
\begin{equation}
\{X_{M},A_{N}\}=\frac{\partial A_{N}}{\partial P_{M}}  \tag{3.7a}
\end{equation}
\begin{equation}
\{P_{M},A_{N}\}=-\frac{\partial A_{N}}{\partial X_{M}}  \tag{3.7b}
\end{equation}
\begin{equation}
\{A_{M},A_{N}\}=0  \tag{3.7c}
\end{equation}
Brackets (3.7a) and (3.7b) are the usual Poisson brackets for a vector
function $A_{M}(X,P).$ Bracket (3.7c) is imposed as an initial simplifying
restriction on the possible functional forms of the gauge field $A_{M}(X,P).$
It is a restriction to Abelian gauge fields.

Computing the algebra of constraints (3.4)-(3.6) using the brackets (2.9)
and (3.7) we obtain the equations 
\begin{equation*}
\{\phi _{1},\phi _{2}\}=-2\phi _{1}+(P^{M}-A^{M})\frac{\partial }{\partial
X^{M}}(X.A)-2(P-A).A
\end{equation*}
\begin{equation*}
-X^{M}\frac{\partial }{\partial X^{M}}[(P-A).A]-X^{M}\frac{\partial }{%
\partial X^{M}}(\frac{1}{2}A^{2})
\end{equation*}
\begin{equation}
+(P^{M}-A^{M})\frac{\partial }{\partial P_{M}}[(P-A).A]+(P^{M}-A^{M})\frac{%
\partial }{\partial P_{M}}(\frac{1}{2}A^{2})  \tag{3.8a}
\end{equation}
\begin{equation*}
\{\phi _{1},\phi _{3}\}=-\phi _{2}+X^{M}\frac{\partial }{\partial P_{M}}[%
(P-A).A]
\end{equation*}
\begin{equation}
-X.A+X^{M}\frac{\partial }{\partial P_{M}}(\frac{1}{2}A^{2})  \tag{3.8b}
\end{equation}
\begin{equation}
\{\phi _{2},\phi _{3}\}=-2\phi _{3}+X^{M}\frac{\partial }{\partial P_{M}}%
(X.A)  \tag{3.8c}
\end{equation}
Equations (3.8)\ exactly reproduce the linear gauge algebra (2.10) when the
conditions 
\begin{equation}
X.A=0  \tag{3.9a}
\end{equation}
\begin{equation}
(P-A).A=0  \tag{3.9b}
\end{equation}
\begin{equation}
\frac{1}{2}A^{2}=0  \tag{3.9c}
\end{equation}
hold. Condition (3.9a) is the first of Dirac's subsidiary conditions [7,8]
on the gauge field. Condition (3.9b) is not an independent condition. It is
the canonical conjugate to condition (3.9a), but incorporating the minimal
coupling prescription 
\begin{equation}
P_{M}\rightarrow P_{M}-A_{M}  \tag{3.10}
\end{equation}
to gauge fields. Condition (3.9c) implies that the canonical vector field $%
A_{M}(X,P)$ is a massless gauge field, describing infinite-range
interactions. When conditions (3.9) hold, the linear gauge algebra (2.10) is
reproduced by constraints (3.4)-(3.6). Thus, when (3.9) holds, the only
possible space-time metric associated with constraints (3.4)-(3.6) giving a
non-trivial surface and avoiding the ghost problem is the flat metric with
two timelike dimensions.

The interacting Hamiltonian (3.3) will be invariant under transformations
(2.13) when the gauge field effectively transforms as 
\begin{equation}
A_{M}\rightarrow \tilde{A}_{M}=\exp \{-\beta \}A_{M}  \tag{3.11}
\end{equation}
which is consistent with transformation (2.13b) and with the minimal
coupling prescription (3.10). Choosing now $\beta =\phi _{1}=\frac{1}{2}%
(P-A)^{2}\approx 0$, and performing the same steps as in the free theory, we
obtain the brackets 
\begin{equation}
\{P_{M},P_{N}\}=0  \tag{3.12a}
\end{equation}
\begin{equation*}
\{X_{M},P_{N}\}=\delta _{MN}-P_{M}P_{N}+P_{N}\frac{\partial }{\partial P_{M}%
}[(P-A).A)]
\end{equation*}
\begin{equation}
+P_{N}\frac{\partial }{\partial P_{M}}(\frac{1}{2}A^{2})-X_{M}\frac{\partial 
}{\partial X_{N}}[(P-A).A]-X_{M}\frac{\partial }{\partial X_{N}}(\frac{1}{2}%
A^{2})  \tag{3.12b}
\end{equation}
\begin{equation*}
\{X_{M},X_{N}\}=-(X_{M}P_{N}-X_{N}P_{M})+X_{M}\frac{\partial }{P_{N}}[(P-A).A%
]
\end{equation*}
\begin{equation}
+X_{M}\frac{\partial }{\partial P_{N}}(\frac{1}{2}A^{2})-X_{N}\frac{\partial 
}{\partial P_{M}}[(P-A).A]-X_{N}\frac{\partial }{\partial P_{N}}(\frac{1}{2}%
A^{2})  \tag{3.12c}
\end{equation}
As can be verified, the above brackets reduce to the Snyder brackets (2.17)
when conditions (3.9) hold. In other words, when conditions (3.9) are valid,
the Lorentz $SO(d,2)$ generator for the two-time physics model with a
bi-local gauge field $A_{M}(X,P)$ described by action (3.2) is identical to $%
L_{MN}$ in (2.11).

Up to now we have proved that, when conditions (3.9) are valid, action (3.2)
has a gauge algebra identical to the linear algebra (2.10) and a Lorentz
generator identical to (2.11). To complete the proof of the consistency of
action (3.2), we must verify the gauge invariance of $L_{MN}$ under gauge
transformations generated by constraints (3.4)-(3.6). Using brackets (2.9)
and (3.7) we find the equations 
\begin{equation*}
\{L_{MN},\phi _{1}\}=X_{M}\frac{\partial }{\partial X_{N}}[(P-A).A]+X_{M}%
\frac{\partial }{\partial X_{N}}(\frac{1}{2}A^{2})
\end{equation*}
\begin{equation*}
-X_{N}\frac{\partial }{\partial X_{M}}[(P-A).A]-X_{N}\frac{\partial }{%
\partial X_{M}}(\frac{1}{2}A^{2})+P_{M}\frac{\partial }{\partial P_{N}}[%
(P-A).A]
\end{equation*}
\begin{equation}
+P_{M}\frac{\partial }{\partial P_{N}}(\frac{1}{2}A^{2})-P_{N}\frac{\partial 
}{\partial P_{M}}[(P-A).A]-P_{N}\frac{\partial }{\partial P_{M}}(\frac{1}{2}%
A^{2})  \tag{3.13a}
\end{equation}
\begin{equation*}
\{L_{MN},\phi _{2}\}=X_{M}\frac{\partial }{\partial X_{N}}(X.A)-X_{N}\frac{%
\partial }{\partial X_{M}}(X.A)
\end{equation*}
\begin{equation}
+P_{M}\frac{\partial }{\partial P_{N}}(X.A)-P_{N}\frac{\partial }{\partial
P_{M}}(X.A)  \tag{3.13b}
\end{equation}
\begin{equation}
\{L_{MN},\phi _{3}\}=0  \tag{3.13c}
\end{equation}
We see from the above equations that $L_{MN}$ is gauge invariant, $%
\{L_{MN},\phi _{i}\}=0$, when conditions (3.9) are valid. Action (3.2),
complemented with the extended Dirac's subsidiary conditions (3.9), gives
therefore a consistent description of two-time physics with background
canonical gauge fields in a phase space with the Poisson brackets (2.9) and
(3.7)

Action (3.2) can also be used to describe two-time physics with background
canonical gauge fields in a phase space with Snyder brackets (2.17) together
with brackets (3.7). This can be seen as follows. First, if we compute the
algebra of constraints (3.4)-(3.6) using these brackets we will find that
the resulting gauge algebra reduces to the nonlinear algebra (2.18) when
conditions (3.9) hold. Second, as we saw in this section, the Lorentz
generator for action (3.2) is identical to $L_{MN}$ when conditions (3.9)
hold. Third, if we compute the brackets $\{L_{MN},\phi _{i}\}$ with $\phi
_{i}$ given by (3.4)-(3.6), using these brackets we will find that $L_{MN}$
is gauge invariant when conditions (3.9) hold. The conclusion of this is
that Dirac's conditions (3.9) are also the necessary subsidiary conditions
for the consistency of the formalism in a phase space with Snyder brackets
(2.17) together with brackets (3.7).

\section{Concluding remarks}

In this work we presented two distinct ways to study two-time physics with
background Abelian canonical gauge fields. In the first way, which will
correspond to a commutative $d+2$ dimensional space-time in the quantized
theory, the fundamental brackets are the Poisson brackets (2.9) together
with (3.7), the $SO(d,2)$ Lorentz generator $L_{MN}$ is given by (2.11), and
the gauge algebra is the linear algebra (2.10). In the second way, which
will correspond to a noncommutative $d+2$ dimensional space-time in the
quantized theory, the fundamental brackets are the Snyder brackets (2.17)
together with (3.7), the $SO(d,2)$ Lorentz generator is the same $L_{MN}$
given by (2.11), and the gauge algebra is the nonlinear algebra (2.18). The
equivalence or not of these two approaches is a subject for future
investigations.

As a final observation, notice that the subsidiary conditions (3.9) appear
in equations (3.8), (3.12) and (3.13) in derivatives with respect to $X_{M}$
and with respect to $P_{M}$. Therefore conditions (3.9), and the formalism
we presented above, remain valid in the more restrictive case when $%
A_{M}=A_{M}(X)$. This case is useful to make contact with the one-time
dynamics, but for formal theoretical investigations in two-time physics the
case with $A_{M}=A_{M}(X,P)$ seems to be more reliable.

\noindent

\end{document}